\documentclass[runningheads,fleqn]{svmult}
\usepackage{makeidx}   
\usepackage{graphicx}  
\usepackage{subeqnar}  
\usepackage{multicol}  
\usepackage{taphys}    
\makeindex             
\newcommand{\text}[1]{\mbox{#1}}
\newcommand{\w}{\omega}
\newcommand{\PP}{\tilde{P}_x}

\newcommand{\sign}{\text{sign}}
\newcommand{\EWERT}[1]{\left\langle #1 \right\rangle}

\begin{document}
\title*{Transport in Quasi One-Dimensional Systems}
\titlerunning{Transport in Quasi One-Dimensional Systems}
%
\author{Achim Rosch}
\authorrunning{A. Rosch}
%
%
\institute{Institut f\"ur Theorie der Kondensierten Materie, Universit\"at Karlsruhe, D-76128 Karlsruhe, Germany}

\maketitle              

\begin{abstract}
  The interplay of Umklapp scattering from a periodic potential and
  other scattering processes determine the conductivity of (quasi)
  one-dimensional metals.  We show that the transport at finite
  temperature is qualitatively and quantitatively strongly influenced
  by a number of approximate conservation laws.  Typically, not the
  strongest but the second strongest scattering mechanism determines
  the dc-conductivity. We discuss the optical conductivity both of
  strongly anisotropic, quasi one-dimensional Fermi liquids and of
  Luttinger liquids.
\end{abstract}

Thanks to the miracles of bosonization, conformal field theory, Bethe
ansatz and renormalization group, the theoretical description of
one-dimensional systems is one of the best developed areas in the
theory of strongly correlated systems \cite{review}. The enormous
theoretical advance is contrasted by failure to understand certain
experimental quasi one-dimensional systems like the Bechgaard salts
even on a qualitative level  \cite{bechgaards}. 
In recent years, carbon nanotubes have
proven to be an almost ideal one-dimensional system where theoretical
concepts can be tested in some detail.

Astonishingly, the theory of finite temperature transport in clean
one-dimensional (1d) systems is not developed very far. The technical
reason for this is that in a pure Luttinger-liquid the conductivity is
infinite and $\sigma$ is therefore a singular function of irrelevant
operators and naive perturbation theory can be problematic and
misleading. In Ref. \cite{PRL}, we argued that main features of the
optical conductivity $\sigma(\w,T)$ can be understood from the
analysis of certain approximate conservation laws and proposed to
calculate $\sigma$ from a hydrodynamic theory of the corresponding
slowly decaying modes. The main result is, that in many experimentally
relevant situations, the strongest scattering process alone cannot
lead to a relaxation of the current due to some conservation law.  As
a consequence, temperature dependence of the dc-conductivity is often
determined by the second strongest scattering processes.  This point of view has
been tested numerically in \cite{garst}.  We will first discuss in
some detail the role of approximately conserved quantities for a quasi
one-dimensional Fermi liquid, afterwards we will shortly review the
results for Umklapp scattering in a Luttinger liquid and discuss how
the results generalize for other scattering mechanisms.

\section{Transport in an Anisotropic Fermi Liquid}

Before we investigate the transport properties of a Luttinger liquid,
we study the role of Umklapp scattering in an anisotropic Fermi
liquid. The arguments given here will be published in a more extended
version in Ref.~\cite{tobe}.

\begin{figure}
\sidecaption
\includegraphics[width=0.4 \linewidth]{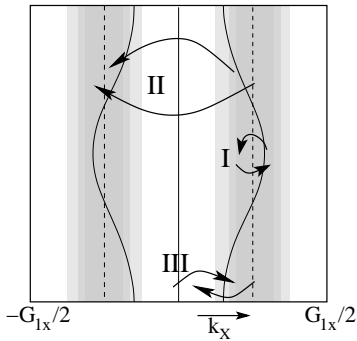}
\caption{Fermi surface of an anisotropic metal. Both ``forward'' (I)
  and ``Umklapp'' (II) scattering processes do not lead to an decay of
  the pseudo momentum $\PP$ as long as the momenta are within the
  shaded area. The scattering event III conserves momentum but leads
  to a decay of $\PP$. It is exponentially suppressed even at moderate
  temperatures as it involves quasi-particles far away from the Fermi
  surface \label{figFS} }\end{figure}

\subsection{Pseudo Momentum Conservation Close to Half Filling}
\label{pseudoFermiLiquid}
We consider an anisotropic metal close to half filling with a clearly
defined most-conducting axis in $x$-direction.  It is assumed that two
well defined Fermi sheets perpendicular to this axis exist (see
Fig.~\ref{figFS}). The curvature of those sheets is not required to be
extremely small, for our argument it is sufficient it is less than
e.g. a quarter of the width of the Brillouin zone (i.e.  within the
shaded area of Fig.~\ref{figFS}). We consider a rather arbitrary
lattice, assuming only the existence of a translation vector
$\vec{a}_1$ of the underlying lattice in x-direction.

Umklapp processes lead to a decay of any macroscopic momentum.  For
example, close to $1/2$ filling a low-energy Umklapp processes (type
II in Fig.~\ref{figFS}) with a momentum transfer $G_x$ relaxes the
momentum in $x$ direction, where $\vec{G}$ is a reciprocal lattice vector. 
However, a pseudo momentum $\PP$ can be
defined, which is conserved by two-particle scattering processes close
to the Fermi surface as we will show in the following
\begin{eqnarray}
\PP=\sum_{\vec{k}\sigma} \delta k_x  c^\dagger_{\vec{k}\sigma}
c_{\vec{k}\sigma}, \quad \text{with } \ \delta k_x=k_x-\frac{G_{1x}}{4} 
\sign[k_x].\label{defineP}
\end{eqnarray}
The pseudo momentum $\delta k_x$ is measured with respect to the line $k_x=\pm
\frac{G_{1x}}{4}$ (dashed line in Fig.~\ref{figFS}).  Here, we
concentrate on systems close to half filling -- for a quasi 1d system
close to a different commensurate filling, other pseudo-momenta are
more relevant \cite{PRL}.

To check to what extent $\PP$ is conserved, we calculate the
commutator of $\PP$ with a generic 2-particle scattering term:

\begin{eqnarray}
H_2&=&\sum_{\text{1. BZ}} c^\dagger_{\vec{k}_1} c^\dagger_{\vec{k}_2}
 c_{\vec{k}'_2} c_{\vec{k}'_1}  V_{\vec{k}_1 \vec{k}_2,\vec{k}'_1 \vec{k}'_2} \sum_{\vec{G}_{\vec{n}}} \delta(\vec{k}_1+\vec{k}_2-
\vec{k}'_1-\vec{k}'_2-\vec{G}_{\vec{n}}) \nonumber \\[0pt]
 [\PP,H_2]&=&\sum_{\text{1. BZ}} c^\dagger_{\vec{k}_1} c^\dagger_{\vec{k}_2}
 c_{\vec{k}'_2} c_{\vec{k}'_1}  V_{\vec{k}_1 \vec{k}_2,\vec{k}'_1 \vec{k}'_2}
(\delta k_{1x}+\delta k_{2x}-\delta k'_{1x}-\delta k'_{2x}) \nonumber \\
&& \times 
\sum_{\vec{G_{\vec{n}}}} \delta(\vec{k}_1+\vec{k}_2-\vec{k}'_1-\vec{k}'_2
-\vec{G}_{\vec{n}} )
\label{comm}
\end{eqnarray}
where the $\vec{G}_n$ are reciprocal lattice vectors.
  It is easy to check, that all those terms on the right-hand side of
  (\ref{comm}) vanish if all four momenta are in the shaded
  region of Fig,~\ref{figFS}, i.e.  $|\delta k_{1x}|$, $|\delta
  k_{2x}|$, $|\delta k'_{1x}|$, $|\delta k'_{2x}|< G_{1x}/4$.  For ``forward
  scattering'' processes of type I in Fig,~\ref{figFS}, $k_{1x}+
  k_{2x}-k'_{1x}-k'_{2x}=0$,  both momentum $P_x$ and pseudo momentum
  $\PP$ are conserved as for all 4 momenta have the same sign.  While
  ``Umklapp'' processes of type II pick up a lattice momentum $G_x$,
  this is exactly compensated by the fact that two electrons are
  moving from the right to the left Fermi surface due to the term
  $\sign[k_x] G_{1x}/4$ in the definition of $\delta k_x$
  (\ref{defineP}) and $\PP$ is again conserved. The pseudo momentum
  can only decay by high-energy processes far from the Fermi surface,
  e.g. III in Fig.~\ref{figFS}. In a two-particle scattering event
  (e.g. in 2nd order perturbation theory) such a scattering process is
  exponentially suppressed even at moderate temperatures.
  
  In high orders of perturbation theory, however, low-energy
  contribution can result from these high-energy processes due to
  virtual excitations. Or to put it in the language of renormalization
  group: $N$-particle interactions are generated. Will they relax
  $\PP$?  From the definition of the pseudo momentum it is clear that
  $\PP$ is conserved modulo $G_{1x}/2$.  Therefore any $N$-particle
  scattering event will change $\PP$ by a multiple of $G_{1x}/2$.
  Accordingly, a relaxation of $\PP$ is not possible if all $2N$
  pseudo momenta $\delta k_{ix}$ involved in the scattering process
  are smaller than $\frac{G_{1x}}{4 N}$.  At low $T$, a decay of $\PP$
  by $N$-particle collision can only happen for
\begin{eqnarray}\label{minN}
N>\frac{G_{1x}/4}{ \max |\delta k_{Fx}|}
\end{eqnarray}
where $\max |\delta k_{Fx}|$ is the maximal distance of the Fermi
surface from the plane $k_x=\pm G_{1x}/4$ (dashed line in
Fig.~\ref{figFS}). 
At sufficiently high temperatures, the broadening of the Fermi-surface
and the thermal excitation of states with higher energy will favor
decay channels of the pseudo momentum with smaller $N$ (this effect
can crudely be described by adding $T/v_F$ to $\max |\delta k_{Fx}|$
in (\ref{minN})).

The temperature dependence of the decay-rate $\Gamma_{\PP}$ of $\PP$
at low $T$ in the Fermi liquid regime is determined by the usual
phase-space arguments: a particle of energy $\w\sim T$ decays in
$2N-1$ particle and hole excitation, one of the energies is fixed by
energy conservation, the remaining $2 N-2$ energies each have a
phase-space of order $\w$ and therefore
\begin{eqnarray}
\Gamma_{\PP} \propto T^{2 N-2}\label{gammaP}
\end{eqnarray}
where the integer $N$ is the smallest value consistent with
(\ref{minN}).  The prefactor in (\ref{gammaP}) depends in a rather
delicate way on the strength and range of the interaction, the
screening and the band-curvature. Note, that a {\em local}
$N$-particle interaction will give no contribution for $N>2$ due to
the Pauli-principle. Therefore the scattering rate is strongly
suppressed for weakly coupled chains with well-screened interactions.
Furthermore, additional logarithmic temperature dependences of the
scattering vertices are expected even in the Fermi-liquid regime as it is
well known from Fermi liquid theory.

We want to stress that the analysis given above is valid for
interactions of arbitrary strength as long as a Fermi liquid
description is possible (For strong interaction, one should, however,
consider the pseudo momentum of quasi-particles which slightly
differs from the pseudo momentum of the bare electrons, defined above).

\subsection{Pseudo Momentum and Conductivity}\label{conduct.section}
In the preceding section we have established that the pseudo momentum
$\PP$ will decay very slowly in a quasi one-dimensional metal.  How
does this influence the optical conductivity $\sigma(\w)$? We will
argue in the following, that the pseudo momentum conservation leads to
well defined peak in the optical conductivity at zero frequency and we
will show how its weight can be calculated reliably. Furthermore, the
$T$-dependence of the dc-conductivity is primarily given by the decay
rate of $\PP$ (\ref{gammaP}) at sufficiently low temperatures.

\begin{figure}
\includegraphics[width=0.9 \linewidth]{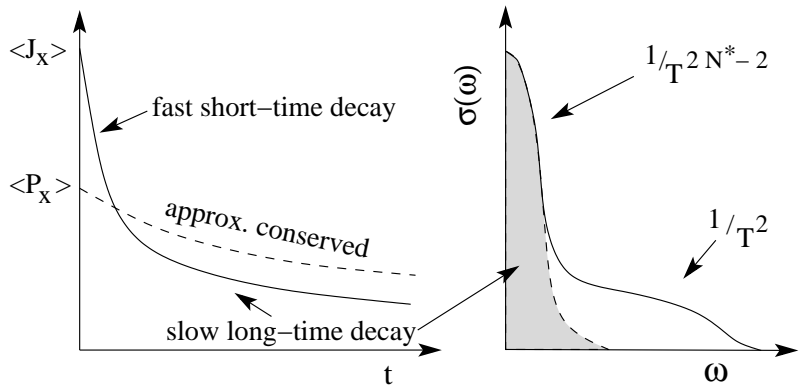}
\caption[]{In a Gedankenexperiment, at time $t=0$ a state with current 
  $\langle J_x \rangle$ is prepared. $J_x$ decays rapidly while the
  pseudo-momentum $\PP$ is approximately conserved and has a much
  lower decay rate $\Gamma_{\PP}$. In the conductivity $\sigma(\w)$
  the approximate conservation of $\PP$ leads to a low-frequency peak
  in the optical conductivity characterized by the decay-rate
  $\Gamma_{\PP}$.  (\ref{gammaP}), here assuming that $N^*$ is the
  smallest integer consistent with (\ref{minN}): a scattering process
  involving $N^*$ particles can relax $\PP$.  The width is given by
  $\Gamma_{\PP}$ (\ref{gammaP}), the weight by (\ref{drude})
\label{figJdecay} }
\end{figure}

We will first try to derive our results using rather simple
hand-waving arguments which will be substantiated by some rigorous
results derived many years ago \cite{mazur,suzuki}. We consider the following Gedankenexperiment: at time $t=0$
we prepare a state with a finite current $\langle J_x(t=0) \rangle>0$.
As the current is not conserved, it will decay rather fast by
two-particle collisions ($\Gamma_{J_x}\propto T^2$ at low $T$).
Typically, the initial state with finite current will also have a
finite pseudo momentum $\langle \PP(t=0) \rangle$ which will decay
much slower than the current; $\Gamma_{\PP}\ll \Gamma_{J_x}$. The
important point is now to realize that any state with finite pseudo
momentum will typically carry a finite current $\langle J_x \rangle=
J(\langle \PP \rangle)$. Accordingly, a finite fraction of $J_x$ will
not decay with the fast rate $\Gamma_{J_x}$ but with the much smaller rate
$\Gamma_{\PP}$ as is shown schematically in Fig.~\ref{figJdecay}.

How large is the fraction of the current which decays slowly? To
answer this question, we have to consider two separate questions:
first, how large is $\langle \PP(t=0)\rangle$, and second, how much
current does a state with finite $\PP$ carry?  At $t=0$ one obtains in
linear response theory the ratio of the expectation values of the
operators $\langle \PP(t=0)\rangle / \langle J_x(t=0)\rangle
=\chi_{\PP J_x}/\chi_{J_x J_x}$ in terms of the corresponding
susceptibilities (defined as usual).  To answer the second question, we
consider a situation where a field conjugated to $\PP$ is applied and find 
in analogy to the argument above $\langle J_x(t\to \infty)\rangle / 
\langle \PP(t\to \infty)\rangle
=\chi_{J_x \PP}/\chi_{\PP \PP}$.
Therefore, we expect (we will discuss later under what condition this
statement is rigorously true) that 
the low-energy peak in the optical conductivity carries a
fraction 
\begin{eqnarray}\label{drudeRel}
\frac{D}{D_0}=\frac{\chi_{J_x \PP}^2}{\chi_{\PP \PP}\chi_{J_x J_x} }
\end{eqnarray}
of the total weight $ \pi \chi_{J_x J_x}=2 \pi D_0=\pi \frac{n
  e^2}{m}$ with $\frac{n}{m}=\sum_{\vec{k \sigma}} \frac{\partial^2
  \epsilon_k}{\partial k^2} \langle c^\dagger_{\vec{k}\sigma}
c_{\vec{k}\sigma}\rangle $. Note, that due to the Cauchy-Schwartz
inequality the relative weight $D/D_0$ is always smaller than 1 as it
has to be.  For lower and lower temperature the peak in $\sigma(\w)$
gets sharper. According to (\ref{drudeRel}), the total weight of the
peak is
\begin{eqnarray}\label{drude}
D=\frac{1}{2} \frac{\chi_{J_x \PP}^2}{\chi_{\PP \PP}}.
\end{eqnarray}
$D$ should not be confused with the zero-temperature Drude weight as
we have not included frequencies of the order of the short-time decay
rate $\Gamma_{J_x}\propto T^2$ in the definition of its weight.

\subsection{Rigorous Results}

It is obvious that the weight of the low-frequency peak in
$\sigma(\w)$ (shaded area in Fig.~\ref{figJdecay}) can rigorously be
extracted only in the limit $\Gamma_{\PP}/\Gamma_{J_x} \to 0$, i.e. in
a situation where $\PP$ is exactly conserved and $D$ is really a
finite-temperature Drude weight. In this situation, Mazur \cite{mazur}
has derived long ago an exact inequality for certain correlation
functions which in our context reads
\begin{eqnarray}\label{mazurInequ}
D\ge \frac{1}{2} \sum_{n=1}^M \frac{\chi_{J_x Q_n}^2}{\chi_{Q_n Q_n}}
\end{eqnarray}
It is valid if all $Q_n$ are conserved and  $\chi_{Q_n Q_m}=0$ for $n \ne m$.
The importance of Mazur's inequality has been recently emphasized  by Zotos
{\it et al.} \cite{zotos2} (see below).

Furthermore, Suzuki \cite{suzuki} showed that the inequality in
(\ref{mazurInequ}) can be replaced by an equality if the sum includes
{\em all} conservation laws! If therefore $\PP$ is the {\em only}
(approximately) conserved quantity in the system with a finite overlap
to the current $\chi_{J_x Q}\ne 0$ (i.e. if $\PP$ is the slowest
``current-like'' mode in the system and $\Gamma_{\PP}$ the smallest
decay rate) then (\ref{drude}) is exact as was tested numerically for
a simple model in \cite{garst}.

\subsection{Low-Frequency Weight in Fermi Liquid Theory}

$\chi_{\PP \PP}$ can easily be calculated at low $T$ within Fermi
liquid theory following standard text books \cite{fermi}.  The result will
in general depend on the details of the momentum dependence of the
effective interactions and the band-structure. We assume a quasi 1D
system with a Fermi velocity $v_F^*=k_F/m^*$ parallel to the most
conducting axis and -- for simplicity -- completely local interactions
characterized by two Fermi liquid parameter $F_{++}$ and $F_{+-}$ in
the spin-singlet channel to describe the interactions of two density
excitations $\delta n_{\vec{k}}$ on the same Fermi sheet or on
different sheets, respectively. With $F_m=F_{++}-F_{+-}$, the relative 
weight of the low frequency peak in the optical conductivity for low $T$
is given by
\begin{eqnarray} \label{drudeFL}
\frac{D}{D_0}\approx \frac{m}{m^*} \left(\frac{ 
\EWERT{(\delta k_x)^2}_{\text{FS}}}
{ \EWERT{\delta k_x}^2_{\text{FS}}}-\frac{F_m}{1+F_m}\right)^{-1}.
\end{eqnarray}
$\EWERT{\dots}_{\text{FS}}$ is defined as an average over the Fermi
sheet, for example $\EWERT{\delta k_x}_{\text{FS}}=\int \!\!\! \int d
k_y d k_z (k_F^x-\frac{G_x}{4})/(\int \!\!\! \int d k_y d k_z)$, where
$k_F^x=k_F^x(k_y,k_z)$ is the x-component of the Fermi momentum on the
right Fermi sheet.  Note that due to Luttinger's theorem $\Delta n=2
\EWERT{\delta k_x}_{\text{FS}}/(a_y a_z \pi)$, where $\Delta n$ is the
deviation of the  electron-density from half filling.

If the interactions are sufficiently weak so that no 
 phase transition is induced, the low-frequency weight $D$
vanishes close to half filling with
\begin{eqnarray}\label{drudeFLhalf}
\frac{D}{D_0} \sim \frac{m}{m^*} \left(\frac{\epsilon_F^*}{t_\perp^*}\right)^2
 \left(\frac{\Delta n}{n}\right)^2.
\end{eqnarray}
where $\epsilon_F^*=k_F v_F^*$ is the renormalized Fermi energy.  We
expect that the low-frequency weight $D$ decreases with increasing
temperature, mainly due to the thermal broadening of $\EWERT{(\delta
  k_x)^2}_{\text{FS}}$. Leading finite-$T$ corrections to
(\ref{drudeFL}) or (\ref{drudeFLhalf}) are of order
$(T/\epsilon_F^*)^2$.

\section{Luttinger Liquids}

The physics of approximate conservation laws and their influence on
transport discussed in the previous section is even more important for
exactly one-dimensional systems.

Surprisingly, little is known about the conductivity $\sigma(T)$ of an
ideal one-dimensional wire in the presence of Um\-klapp scattering
induced by a periodic potential.  For a long time, there has not even
an agreement whether $\sigma$ is finite or infinite at $T>0$ for
generic systems
\cite{giamarchi,millis,fujimoto,zotos,zotos2,everybody}. For example,
T. Giamarchi \cite{giamarchi,millis} found a finite conductivity for
$T>0$ within a certain perturbation theory.  However, using a
Luther-Emery transformation of the dominant Umklapp process, he
concluded that the conductivity is actually infinite \cite{giamarchi}.
Furthermore, Castella, Zotos and others \cite{fujimoto,zotos,zotos2}
were able to calculate the Drude weight for a number of exactly
solvable, integrable lattice models and they also found an infinite
conductivity for $T>0$. From our discussion in \ref{conduct.section}
this is not too surprising as any integrable model is characterized by
an infinite number of conservation laws -- this point of view has been
emphasized by Zotos {\it et al.} \cite{zotos2}, who proposed to use
(\ref{mazurInequ}) to obtain an estimate for the Drude weight in these
models.  On general grounds, however, one would expect a finite
conductivity in any {\em generic} lattice model.  Most of the above
discussed discrepancies can be resolved from an analysis of the
approximate conservation laws of the appropriate low-energy theories
as has been realized recently by the author and N. Andrei \cite{PRL}.
Within a certain hydrodynamic theory of the approximately conserved
quantities, the main characteristics of the optical conductivity can
be calculated reliably.

\subsection{Pseudo Momenta}

\begin{figure}
\sidecaption
\includegraphics[width=0.38 \linewidth]{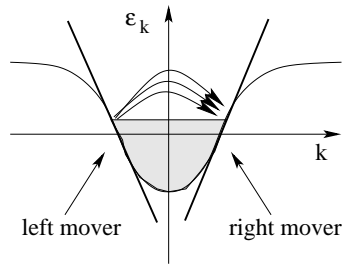}
\caption{\label{umklapp.fig} In $d=1$, a metal is characterized by two Fermi points and the fermionic excitations can be separated in left- and right-movers.
  Certain Umklapp processes (here for $n=3$) can scatter electrons
  from the left to the right Fermi surface and vice versa }\end{figure}

The topology of the Fermi surface of a $1d$ metal determines its
  low-energy excitations. Two well defined
 Fermi-points exist at momenta $k=\pm k_F$,  allowing us to define
 left and right moving
excitations, to be described by $\Psi_{L/R,\sigma=\uparrow 
\downarrow}$ (see Fig.~\ref{umklapp.fig}). We shall include in the fields 
 momentum modes extending to the edge of the Brillouin zone,
usually omitted in treatments that concentrate
on physics very close to the Fermi-surface.

 The
Hamiltonian, including high energy processes, is
\begin{eqnarray}
H=H_{LL}+H_{\text{irr}}+\sum_{n,m}^\infty H_{n,m}^U.
\end{eqnarray}
 $H_{LL}$ is the well-known Luttinger liquid Hamiltonian 
capturing the low energy behavior \cite{review},
\begin{eqnarray}
H_{LL}&=&v_F \int  \left(\Psi_{L
\sigma}^\dagger i \partial_x \Psi_{L \sigma}- \Psi_{R\sigma}^\dagger
i \partial_x \Psi_{R\sigma} \right)  + g \int  \rho(x)^2  \nonumber
\\ &=& \frac{1}{2} \int \frac{dx}{2 \pi}
\sum_{\nu=\sigma,\rho}  v_\nu\left[ K_\nu (\partial_x
\theta_\nu)^2+\frac{1}{K_\nu} (\partial_x \phi_\nu)^2 \right] \nonumber
\end{eqnarray}
$v_F$ is the  Fermi velocity, $g>0$
 measures the strength of interactions, $\rho=\rho_L+\rho_R$
 is the sum of the left and right moving electron densities. 
In the  second line we wrote the
 bosonized \cite{review}
version of the Hamiltonian. Here 
 $v_\sigma$, $v_\rho$ are the spin and charge velocities, and  the
 interactions determine the Luttinger parameters $K_\nu$ with
 $v_\nu K_\nu=v_F$,
 $v_\rho/K_\rho=v_F+g/\pi $, $v_\sigma/K_\sigma=v_F-g/\pi$.

 The high energy processes are captured in the subsequent terms which
 are formally irrelevant at low energies (we
 consider only systems away from a Mott transition, i.e. away from half
 filling). Some of them, however, determine the low-frequency behavior
 of the conductivity at any finite $T$, since they induce the
 decay of the conserved modes of $H_{LL}$ (they are ``dangerously
 irrelevant'').  We classify these irrelevant terms with the help of
 two operators which will play the central role in our discussion. The
 first one is the translation operator $P_T$ of the right- and
 left-moving fields, the second one, $J_0 =N_R-N_L$, is the difference
 of the number of right- and left-moving electrons, and is up to
 $v_F$, the charge current of $H_{LL}$:
\begin{eqnarray}\label{PT}
P_T&=&\sum_\sigma \int dx \left[\Psi_{R
\sigma}^\dagger (-i \partial_x) \Psi_{R \sigma}+ \Psi_{L\sigma}^\dagger
(-i \partial_x) \Psi_{L\sigma}\right] \\ \label{J0}
J_0&=&N_R-N_L= \sum_{\sigma} \int dx  \left[ \Psi_{R
\sigma}^\dagger \Psi_{R\sigma}-\Psi_{L\sigma}^\dagger
\Psi_{L\sigma} \right]
\end{eqnarray}
  The linear
combination $P_0=P_T + k_F J_0$ can be identified with the total
momentum of the full Hamiltonian $H$.

We proceed to the classification of the formally irrelevant terms 
in the Hamiltonian. This classification  allows us to select all those 
terms (actually few in number)
that determine the current dynamics.
$H_{\text{irr}}$ includes all terms in $H-H_{LL}$
 which  commute with both $P_T$ and $J_0$, such as
 corrections due to the finite band curvature,
 due to finite-range interactions and similar terms. We will not need
their explicit form.

The Umklapp terms $H^U_{n,m}$ ($n,m=0,1,...$) convert $n$ right-movers
to left-movers (and vice versa) picking up lattice momentum $m 2
\pi/a=m G$, and do not commute with either $P_T$ or $J_0$ (see
Fig.~\ref{umklapp.fig}). Leading terms are of the form,
\begin{eqnarray}\label{HU}
H^U_{0,m}&\approx&  g^U_{0,m} \int  e^{i \Delta k_{0,m} x} (\rho_L+\rho_R)^2 + h.c. \\
H^U_{1,m}&\approx&  g^U_{1,m} \sum_{\sigma}\int  e^{i \Delta k_{1,m} x} \Psi^\dagger_{ R \sigma} \Psi_{ L \sigma} 
\rho_{-\sigma}+h.c. \\
H^U_{2,m}&\approx&
 g^U_{2,m} \int  e^{i \Delta k_{2,m} x} 
\Psi^\dagger_{ R \uparrow}\Psi^\dagger_{ R \downarrow} 
\Psi_{ L \downarrow}\Psi_{ L \uparrow}+h.c. 
\end{eqnarray}
with momentum transfer $
\Delta k_{n,m}= n 2 k_F-m G$.

The important step is to realize that certain pseudo momenta, $\tilde{P}_{nm}$,
defined as linear combinations of $J_0$ and $P_T$
\begin{eqnarray}
\tilde{P}_{nm}= \frac{\Delta k_{nm}}{2 n} J_0 + P_T
\end{eqnarray}
are approximately conserved. For $m=0$, $\tilde{P}_{n0}$ is nothing
but the usual momentum, for $n=2, m=1$ we recover the pseudo momentum
discussed in section~\ref{pseudoFermiLiquid}.

If the Hamiltonian includes only a {\em single} type of Umklapp process
 $H^U_{nm}$, then the pseudo momentum is {\em exactly} conserved
\begin{eqnarray}
\left[H_{LL}+H_{\text{irr}}+ H^U_{nm}, \tilde{P}_{nm} \right]=0.
\end{eqnarray}
even in the presence of band-curvature and similar terms. Therefore we 
can  expect an infinite conductivity in such a model.  At least two independent
Umklapp terms are required to lead to a complete decay of the current
and a finite conductivity (an exception are exactly commensurate
systems, see  \cite{PRL}). 

In any generic lattice model, certainly all types of Umklapp processes
are present or are ``generated'' in the language of renormalization
group.  It is clear that in such a situation, the strongest Umklapp
process will not determine the $T$ dependence of the dc-conductivity
as the associated conservation law prohibits the decay of the current.
Instead, the second strongest Umklapp determines the decay rate of the
current as we have argued in section~\ref{conduct.section}. As the
second strongest Umklapp process is typically far from the Fermi
surface, the conductivity can be very large. For lowest temperatures,
we find in~\cite{PRL} close to a commensurate filling $n\approx M/N$,
where $M$ and $N$ are integers with with $N>2$
\begin{eqnarray}
\sigma(n=\frac{M}{N}+\Delta n,T) \sim
\max\left[(\Delta n)^2 e^{\beta v G/N},T^{-N^2 K_{\rho}}\right]
\end{eqnarray}
while $\sigma(T) \sim \exp[-(T_0/T)^{2/3}]$ at typical incommensurate
fillings.  Details and omitted prefactors can be found in
Ref.~\cite{PRL}.  The calculation of the conductivity in
Ref.~\cite{PRL} is based on the memory matrix formalism
 \cite{forster}. For a certain classical model of charge excitations of
a weakly doped Mott insulator these effects have been calculated
numerically in  \cite{garst}. These numerical calculations have served
as a test for the analytical methods used in  \cite{PRL}.

\subsection{Competition of Scattering Processes and Role of Integrability}

If one analyses various scattering processes in (quasi)
one-dimensional materials, one realizes that many of them can be
characterized by approximate conservation laws. We argued above, that
Umklapp scattering $H^U_{ n,m}$ conserves pseudo-momentum but also
scattering from low-energy phonons or slowly varying potential
fluctuations are characterized by well-known conserved quantities as
is shown in table~\ref{t1}. The consequences can be that e.g. in a
situation with very strong Umklapp scattering the $T$ dependence of
the conductivity is determined by phonons, while in a regime where
phonons dominate, the power-laws associated to Umklapp processes show
up in $\sigma(T)$.

\begin{table}[h]
\caption{\label{t1}
Scattering processes and the associated conservation laws}
\begin{tabular}{l|l}
scattering mechanism & conserved quantity \\
\hline 
\hline
Umklapp $H^U_{ n,m}$ 
& 
pseudo-momentum $\tilde{P}_{{n},{m}}$ 
\\
\hline
acoustic phonons (1D or 3D)  &   ${ N_R-N_L}$
 \\
\hline
long-range disorder(forward scattering) &  ${ N_R-N_L}$ \\
\hline 
short-range impurities & no conservation \\
\hline
\end{tabular}
\end{table}

This analysis applies both to one- and quasi one-dimensional materials
independent of whether the system is better described by a Luttinger
liquid, a Fermi liquid or something else (however, the $T$ dependence
of $\sigma$ will be very different).

Generically, the low-energy behavior of exactly one-dimensional
systems is characterized by an integrable fixed-point Hamiltonian with
an {\em infinite} number of conservation laws.  It is therefore
important to analyze whether these more complicated conservation laws
influence measurable quantities like the conductivity for {\em
  generic} models. For incommensurate systems it is easy to check that
at low temperature it is sufficient to keep track of the
pseudo-momentum conservation as the decay rate of the relevant
$P_{nm}$ is exponentially small in temperature while other more
complicated quantities decay with algebraically small rates. The
situation is more complicated for an exactly commensurate filling as
$\chi_{J \tilde{P}_{n,m}}$ vanishes with exponential precision if
$\Delta k_{nm}=0$  \cite{PRL,tobe}.  In this situation one has to
analyze in more detail the influence of other approximate conservation
laws. Preliminary results suggest that for 1d Mott insulators with a
gap $\Delta$, and for temperatures $\Delta < T < \epsilon_F$, more
complicated conservation laws of the associated fixed point
Hamiltonian (the sine-Gordon model in this case) are indeed important.
This implies, that the optical conductivity will develop a well
defined low-frequency peak as sketched in Fig.~\ref{figJdecay}
associated to slowly decaying modes which have their origin in the
structure of conservation laws of the sine-Gordon model.

\section{Conclusions}

In this paper, we have discussed various approximate conservation laws
which determine the low-frequency conductivity of clean (quasi)
one-dimensional materials. For a large class of situations, the
dominant scattering process leads not to a decay of the current due to
the presence of a slowly decaying mode. In this situation the second
strongest scattering process determines the dc-conductivity. The most
important signature of this type of physics is a well-defined low
frequency peak in the optical conductivity as is shown schematically
in Fig.~\ref{figJdecay}.  The weight of such a peak can be calculated
from (\ref{drude}).  In experiments on Bechgaard salts
 \cite{bechgaards} indeed a well defined low-frequency peak with a
small $1\%$ weight have been found. Presently, it is however not clear
whether one of the conservation laws discussed in this paper is at the
origin of this feature.

I would like to thank M.~Garst, P.~W\"olfle, X.~Zotos and especially
N.~Andrei for many valuable discussions. This work was supported by
the Emmy-Noether program of the DFG.

%

\end{document}